\documentstyle[aps,prl,epsf]{revtex}

\begin{document}
\draft
\title{The Three-Magnon Contribution to the Spin Correlation Function in
Integer-Spin
Antiferromagnetic Chains}
\author{M.D.P. Horton$^1$ and Ian Affleck$^2$}
\address{
$^1$Department of Physics and Astronomy, The University of British Columbia,
Vancouver, B.C., V6T 1Z1, Canada \\
$^2$Department of Physics and Astronomy and Canadian
Institute for Advanced
Research, The University of British Columbia,
Vancouver, B.C., V6T 1Z1, Canada}

\date{\today}

\maketitle

\begin{abstract}
The exact form factor for the O(3) non-linear $\sigma$ model
is used to predict the three-magnon contribution to
the spin correlation function, $S(q,\omega )$, near wavevector $q=\pi$ in an integer spin,
one-dimensional antiferromagnet. The  3-magnon contribution is extremely broad and 
extremely weak; the integrated intensity is  $<2\%$ of the single-magnon contribution. 
\end{abstract}

\pacs{PACS number 75.10.Jm}

The Hamiltonian for the one dimensional Heisenberg antiferromagnet of spin $s$ is
\begin{equation}
 H=J\sum_{i}\vec S_{i}\cdot \vec S_{i+1}. 
\end{equation}
Based upon the large-s limit  we write the spin operators as
\begin{equation}
\vec S_{j} \approx s(-1)^{j}\vec \phi (j)+\vec l(j),
\end{equation}
where $s\vec \phi$ and $\vec l$ represent the staggered and uniform
magnetization of the spin chain. We set the lattice spacing to 1.  The low energy behaviour of this
Hamiltonian can be described by the O(3) non-linear $\sigma$ model~\cite{Affleck1}.
Recently, the exact form factors for this field theory
were calculated~\cite{Karowski,Krillov,Balog}.  The resulting prediction for $S(q,\omega )$ at $q\approx 0$ was
discussed in \cite{Affleck2}.  Here we comment on the prediction at $q\approx \pi$.  
In the continuum approximation, the (zero temperature) 
spin correlation function at $q=\pi + k \approx \pi $  is
\begin{equation}
 S^{ab}(\pi +k,\omega)=s^2\int dx \int dt \exp{i(\omega t-kx)} <\Omega\mid\phi^a(x,t)\phi^b(0,0)\mid\Omega>\equiv 
\delta^{ab}S(\pi +k,\omega ) ,
\end{equation}
where  $|\Omega >$ is the groundstate. 

We insert a complete set of asymptotic states. It is known that the spectrum consists of a triplet of massive magnons.  Thus, the
asymptotic states will be characterized by a spin index and a momentum for
each particle in the state.
\[\mid n> = \mid a_1,p_1;a_2,p_2;\ldots ;a_n,p_n> \]
It is convenient to label the particles' momenta by the  rapidities, $\theta_i$:
\begin{equation}
E_i=\Delta \cosh \theta_i,\ \  p_i=(\Delta /v)\sinh \theta_i\end{equation}
where $v$ is the spin-wave velocity corresponding to the velocity of light in
the quantum field theory and $\Delta$ is the gap corresponding to
the rest mass energy in the quantum field theory. ($v\approx 2.49J$ and $\Delta \approx .4107J$ for the s=1 chain.)  Thus we may write:
\begin{equation}
 S^{ab}(\pi +k,\omega)=s^2\int dx \int dt \exp{i(\omega t-kx)} \sum_{n} \frac{1}{n!}  \prod_{i=1}^{n}  \int \frac{d\theta_{i}}{4\pi}
<\Omega\mid\phi^a(x,t)\mid n><n\mid \phi^b(0,0)\mid\Omega>.\label{complete}
\end{equation}
We use 
\begin{equation}
<\Omega\mid\phi(x,t)\mid n> = <\Omega \mid \phi(0) \mid n> \exp[-i(E_nt-P_nx)],
\end{equation}
where $P_n$ and $E_n$ refer to the total momentum and energy of the state $|n>$, to obtain
\begin{equation} 
 S(\pi +k,\omega )=s^2 \frac{(2\pi)^2}{3} \sum_{a}  \sum_{n} \frac{1}{n!} \prod_{i=1}^{n} \int \frac{d\theta_{i}}{4\pi}
\delta(k-P_n)\delta(\omega-E_n) \mid <\Omega\mid\phi^a(0,0)\mid n> \mid ^2.
\end{equation}

The field is renormalized
\begin{equation}
\Phi^{a}(x)=\frac{1}{\sqrt{Z}}\phi^{a}(x)
\end {equation}
in order that we satisfy the relation $< \Omega \mid \Phi^{a}(0) \mid b,p>
= \delta^{ab}$.  Symmetry arguments guarantee
that only asymptotic states with an odd number of magnons will offer
non-zero matrix elements.

For the one particle contribution,
\begin{equation}
S_1(\pi +k,\omega )=s^2vZ\pi \frac {\delta (\omega -\sqrt{v^2k^2 + \Delta^2})}{\sqrt {v^2k^2 + \Delta^2}}. 
\label{S1omega}\end{equation}
The integrated intensity is,
\begin{equation}
 S_1(q) \equiv \int dw S_1(q,w) = s^2vZ\pi \frac
{1}{\sqrt {v^2k^2 + \Delta^2}}.
\end{equation}
Numerical simulations on the $s=1$ antiferromagnet\cite{Sorensen} indicate that $Z\approx 1.26$.

The three particle contribution can be written:
\begin{eqnarray} 
 S_3(\pi +k,\omega)&=&s^2vZ \frac {\pi}{\sqrt{\omega^2-v^2k^2}} \int_0^\infty \frac{d(\theta_1-\theta_2)
d(\theta_2-\theta_3)}{(4\pi)^2}\delta(\sqrt{\omega^2-v^2k^2}-M(\theta_1,\theta_2,\theta_3))\nonumber \\
&& \frac{1}{3}\sum_{a,a_1,a_2,a_3} \mid <\Omega \mid \Phi^a(0,0) \mid
a_1,p_1;a_2,p_2;a_3,p_3 > \mid ^2,
\label{ff}\end{eqnarray}
where:
\begin{equation}
M\equiv \sqrt{\left( \sum_{i=1}^3E_i\right)^2-v^2
\left( \sum_{i=1}^3p_i\right)^2} \ \ \hbox{and}\ \  \theta_{ij}\equiv \theta_i-\theta_j.
\end{equation}
Remarkably, the 3-particle form factor  has been calculated exactly using the integrability of the non-linear $\sigma$ model\cite{Krillov,Balog}:
\begin{eqnarray}
\frac{1}{3}\sum_{a,a_1,a_2,a_3}
\mid <\Omega \mid \Phi^a(0,0) \mid
a_1,p_1;a_2,p_2;a_3,p_3 > \mid ^2&=&\pi^6|\psi (\theta_1,\theta_2,\theta_3)|^2
[2(\theta_{21}^2+\theta_{32}^2+\theta_{31}^2)+12\pi^2],
\nonumber \\
\psi (\theta_1,\theta_2,\theta_3)&\equiv& \prod_{i>j}\psi(\theta_{ij}), \nonumber \\
\psi(\theta )&\equiv& {\theta-i\pi\over \theta (2\pi i-\theta )}\tanh^2{\theta \over 2}.
\label{exff}\end{eqnarray}
 The resulting integral in Eq. (\ref{ff}) can be easily performed numerically.
The result for $S(\pi ,\omega )$ is shown in Fig. 1.
\begin{figure}
\epsfysize 5cm
\epsfbox[60 500 550 700]{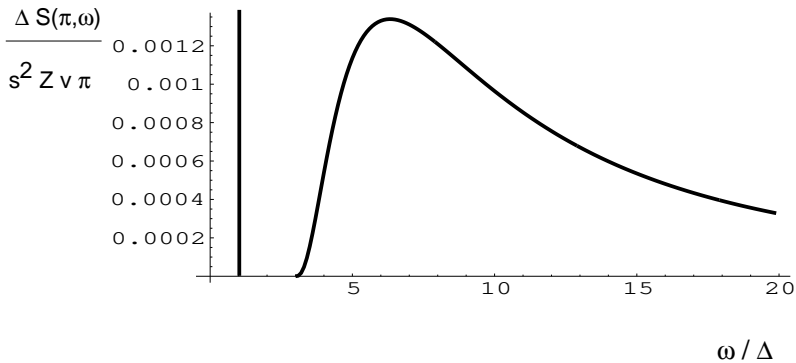}
\caption[]{1-magnon and 3-magnon contribution to spin-correlation
function, $S(\pi ,\omega )$.  The factor of $s^2vZ\pi /\Delta$ in Eq. (\protect\ref{S1omega}) has been divided out so that the peak
 at $\omega =\Delta$ has unit integral.}
\label{fig1}
\end{figure}
The three-magnon contribution vanishes below $3\Delta$. In the limit $\omega \to 3\Delta$, it behaves as:\begin{equation}
S_3(\pi ,\omega )\to s^2vZ\times .01045(\omega -3\Delta )^{3}/\Delta^5. 
\end{equation}
It has a rounded, asymmetric peak at $\omega \approx 6.33\Delta$ then decays at high 
energy as $s^2vZ\times 19.9/\{ \omega^2[\ln (\omega /\Delta )]^{2}\} $.
The integrated intensity of the three particle
contribution at $q=\pi$ is
\begin{equation}
 S_3(\pi ) \approx .0193 S_1(\pi ).
\end{equation}
This 3-particle contribution to $S(q,\omega )$ is very weak and very broad.  It is instructive to
calculate the average frequency of the 3-particle term:
\begin{equation}
\bar \omega_3\equiv {\int d\omega \omega S_3(\pi ,\omega )\over \int d\omega S_3(\pi ,\omega )}\label{av}
\end{equation}
Using the result of \cite{Balog} for the integral in the numerator of Eq. (\ref{av}) we find:
\begin{equation}
\bar \omega_3\approx 75.2\Delta .\end{equation}
This enormous value is explained, to some extent, by the rapid decrease of $S_3(\pi ,\omega )$ at $\omega \to 3\Delta$ 
and its slow drop off  at large
$\omega$, making the integral in the numerator of Eq. (\ref{av}) only logarithmically convergent.  
The contributions to $S$ from still higher numbers of particles have also been considered \cite{Krillov,Balog}.
Indications are that these are still more negligible than this tiny 3-particle term.  

It is clear that this result for $S_3$ cannot be applied completely to the s=1 antiferromagnetic chain.  
In particular, from the numerically determined single magnon dispersion relation, which is quite well fit by:
\begin{equation}
E(p)\approx \sqrt{\Delta^2+v^2\sin^2p},
\end{equation}
it can be seen that
the maximum possible energy of a 3 magnon state with total crystal momentum $\pi$ is only about $17\Delta$.  
$S(\pi ,\bar \omega_3 )$ in the field theory gets significant contributions from  bosons with momenta considerably
larger than $\pi$, the maximum possible in the lattice model.  In fact the relativistic approximation
to the dispersion relation seems to break down significantly for $p>.2\pi$.  We note that, for $\omega <9\Delta$,
only magnons with $p<.2\pi$ contribute to $S_3(\pi ,\omega )$.
  Thus we might hope that $S_3(q,\omega )$
calculated from the field theory is fairly accurate for $\pi -q<.2\pi$ and $\omega <9\Delta$. 
The non-relativisitic corrections to the magnon dispersion relation make $S_3(\pi ,\omega )$ vanish
for $\omega $ greater than about $17\Delta$.  If we only integrate over $S_3(\pi ,\omega )$ up to 
$\omega =17\Delta$ this reduces $S_3(\pi )$ to about $.012 S_1(\pi )$.
However, the non-relativistic corrections may also tend to increase $S_3(\pi ,\omega )$ for
$9\Delta <\omega < 17\Delta$ since they flatten the dispersion relation hence increasing the density
of states.  Clearly the value of $\bar \omega_3$
in the s=1 chain must be less than the maximum possible frequency of about $17\Delta$. 
If we make a rough estimate that $\bar \omega_3\approx 10\Delta$ and $S_3(\pi )\approx .02S_1(\pi )$ then
we can estimate the overall
average frequency (also ignoring 5 and more magnon contributions) as:
\begin{equation}\bar \omega \approx {\int d\omega [S_1(\pi ,\omega )+S_3(\pi ,\omega )]\omega \over 
\int d\omega [S_1(\pi ,\omega )+S_3(\pi ,\omega )]}\approx \Delta + 
{S_3(\pi )\over S_1(\pi )}\bar \omega_3\approx \Delta [1+.02 \times 10]=1.2\Delta .
\label{crude}\end{equation}
This estimate can be compared to the numerically
determined value.  We may use an exact sum rule for the Heisenberg antiferromagnet:
\begin{equation}
\int d \omega \omega S(q,\omega )=-(1/2)<[[H,S^z(q)],S^z(-q)]>=(2/3)|e_0|(1-\cos q),\end{equation}
where $e_0$ is the groundstate energy per site.  Using the numerically determined value of $S(q)$ and $e_0$
gives $\bar \omega (q)$.  This is plotted in Fig. (18) of Ref. \onlinecite{Sorensen} where it is referred
to as $\omega_{\hbox{SMA}}$.  From this figure we see that at $q=\pi$, $\bar \omega \approx 1.2\Delta$ in agreement with our crude estimate of Eq. (\ref{crude}).
This lends some confidence
to  our prediction that the 3-magnon contribution to $S(q,\omega )$ near $q\approx \pi$ is extremely broad
and extremely weak, with $\bar \omega_3$ of order the maximum possible 3-magnon energy, $\approx 10\Delta$
and the relative integrated intensity of order 2\%.  As such, it will be extremely difficult to observe
experimentally.  We note that the 2-particle contribution to $S(q,\omega )$ near $q=0$ is also extremely
difficult to observe \cite{Affleck1,Sorensen,Ma} since it appears to only occur for $q<.2\pi$ and since
$S(q,\omega )\propto q^2$ as $q\to 0$.  Thus experimental applications of the beautiful exact results 
\cite{Zamolodchikov,Karowski,Krillov,Balog} on the
non-linear $\sigma$ model remain elusive.  

If we assume that the 3-particle form factor in Eq. (\ref{exff}) goes to a non-zero constant at
the top of the 3-particle continuum, $\omega \approx 17\Delta$, then it follows from phase space
considerations that $S_3(\pi ,\omega )$ drops discontinously to 0 at that energy.  This is quite
different than the continous vanishing of $S_3$ at the bottom of the 3-particle continum which
is entirely due to the vanishing of the form factor there. Possibly this discontinous drop
at the top of the continuum might be easier to observe experimentally than other features.
However clearly the tiny size of the drop would make even this extremely difficult.

We would like to thank Bill Buyers for interesting us in this problem.  This research was supported in
part by NSERC of Canada.

\end{document}